\documentclass[journal]{IEEEtran}
\usepackage[cmex10]{amsmath}
\usepackage{amsthm}
\usepackage{amssymb}
\usepackage{array}
\usepackage{cite}
\usepackage{hyperref}
\newtheorem{thm}{Theorem}
\newtheorem{cor}[thm]{Corollary}
\newtheorem{lem}[thm]{Lemma}

\theoremstyle{definition}
\newtheorem{defn}[thm]{Definition}
\theoremstyle{remark}
\newtheorem{rem}[thm]{Remark}
\numberwithin{equation}{section}

\begin{document}
\title{Gaussian States Minimize the Output Entropy of the One-Mode Quantum Attenuator}
\author{Giacomo De Palma, Dario Trevisan, Vittorio Giovannetti
\thanks{G. De Palma is with QMATH, Department of Mathematical Sciences, University of Copenhagen, Universitetsparken 5, 2100 Copenhagen, Denmark; NEST, Scuola Normale Superiore, Istituto Nanoscienze-CNR and INFN, I-56126 Pisa, Italy.}
\thanks{D. Trevisan is with Universit\`a degli Studi di Pisa, I-56126 Pisa, Italy.}
\thanks{V. Giovannetti is with NEST, Scuola Normale Superiore and Istituto Nanoscienze-CNR, I-56126 Pisa, Italy.}
\thanks{This paper was presented at the 11th Conference on the Theory of Quantum Computation, Communication, and Cryptography, Berlin (Germany), September 2016; at the Beyond IID in Information Theory 4 Workshop, Barcelona (Spain), July 2016; at the Quantum Roundabout Conference, Nottingham (United Kingdom), July 2016; and at the 13th Central European Quantum Information Processing Workshop, Valtice (Czech Republic), June 2016.}}

\maketitle
\begin{abstract}
We prove that Gaussian thermal input states minimize the output von Neumann entropy of the one-mode Gaussian quantum-limited attenuator for fixed input entropy.
The Gaussian quantum-limited attenuator models the attenuation of an electromagnetic signal in the quantum regime.
The Shannon entropy of an attenuated real-valued classical signal is a simple function of the entropy of the original signal.
A striking consequence of energy quantization is that the output von Neumann entropy of the quantum-limited attenuator is no more a function of the input entropy alone.
The proof starts from the majorization result of De Palma et al., IEEE Trans. Inf. Theory 62, 2895 (2016), and is based on a new isoperimetric inequality.
Our result implies that geometric input probability distributions minimize the output Shannon entropy of the thinning for fixed input entropy.
Moreover, our result opens the way to the multimode generalization, that permits to determine both the triple trade-off region of the Gaussian quantum-limited attenuator and the classical capacity region of the Gaussian degraded quantum broadcast channel.
\end{abstract}

\section{Introduction}
\IEEEPARstart{M}{ost} communication schemes encode the information into pulses of electromagnetic radiation, that is transmitted through metal wires, optical fibers or free space, and is unavoidably affected by signal attenuation.
The maximum achievable communication rate of a channel depends on the minimum noise achievable at its output.
A continuous classical signal can be modeled by a real random variable $X$.
Signal attenuation corresponds to a rescaling $X\mapsto\sqrt{\lambda}\,X$, where $0\leq\lambda\leq1$ is the attenuation coefficient (the power of the signal is proportional to $X^2$ and gets rescaled by $\lambda$).
The noise of a real random variable is quantified by its Shannon differential entropy  $H$ \cite{cover2006elements}.
The Shannon entropy of the rescaled signal is a simple function of the entropy of the original signal \cite{cover2006elements}:
\begin{equation}\label{scaling}
H\left(\sqrt{\lambda}\;X\right)=H\left(X\right)+\ln\sqrt{\lambda}\;.
\end{equation}
This property is ubiquitous in classical information theory.
For example, it lies at the basis of the proof of the Entropy Power Inequality \cite{dembo1991information,gardner2002brunn,shannon2001mathematical,stam1959some,verdu2006simple,rioul2011information,cover2006elements}.

Since the energy carried by an electromagnetic pulse is quantized, quantum effects must be taken into account \cite{gordon1962quantum}.
They become relevant for low-intensity signals, such as for satellite communications, where the receiver can be reached by only few photons for each bit of information \cite{chen2012optical}.
In the quantum regime the role of the classical Shannon entropy is played by the von Neumann entropy \cite{wilde2013quantum,holevo2013quantum} and signal attenuation is modeled by the Gaussian quantum-limited attenuator \cite{chan2006free,braunstein2005quantum,holevo2013quantum,weedbrook2012gaussian,holevo2015gaussian}.

A striking consequence of the quantization of the energy is that the output entropy of the quantum-limited attenuator is not a function of the input entropy alone.
A fundamental problem in quantum communication is then determining the minimum output entropy of the attenuator for fixed input entropy.
Gaussian thermal input states have been conjectured to achieve this minimum output entropy \cite{guha2007classicalproc,guha2007classical,guha2008entropy,guha2008capacity,wilde2012information,wilde2012quantum}.
The first attempt of a proof has been the quantum Entropy Power Inequality (qEPI) \cite{konig2013limits,konig2014entropy,de2014generalization,de2015multimode}, that provides the lower bound
\begin{equation}\label{qEPI}
S\left(\Phi_\lambda\left(\hat{\rho}\right)\right)\geq n\;\ln\left(\lambda\left(e^{\left.S\left(\hat{\rho}\right)\right/n}-1\right)+1\right)
\end{equation}
to the output entropy of the $n$-mode quantum-limited attenuator $\Phi_\lambda$ in terms of the entropy of the input state $\hat{\rho}$.
However, the qEPI \eqref{qEPI} is \emph{not} saturated by thermal Gaussian states, and thus it is not sufficient to prove their conjectured optimality.

Here we prove that Gaussian thermal input states minimize the output entropy of the one-mode quantum-limited attenuator for fixed input entropy (Theorem \ref{thmmain}).
The proof starts from a recent majorization result on one-mode Gaussian quantum channels \cite{de2015passive}, that reduces the problem to input states diagonal in the Fock basis.
The key point of the proof is a new isoperimetric inequality (Theorem \ref{thmiso}), that provides a lower bound to the derivative of the output entropy of the attenuator with respect to the attenuation coefficient.

The restriction of the one-mode quantum-limited attenuator to input states diagonal in the Fock basis is the map acting on discrete classical probability distributions on $\mathbb{N}$ known in the probability literature under the name of thinning \cite{de2015passive}.
The thinning has been introduced by R\'enyi \cite{renyi1956characterization} as a discrete analogue of the rescaling of a continuous real random variable.
The thinning has been involved with this role in discrete versions of the central limit theorem \cite{harremoes2007thinning,yu2009monotonic,harremoes2010thinning}
and of the Entropy Power Inequality \cite{yu2009concavity,johnson2010monotonicity}.
Most of these results require the ad hoc hypothesis of the ultra log-concavity (ULC) of the input state.
In particular, the Restricted Thinned Entropy Power Inequality \cite{johnson2010monotonicity} states that the Poisson input probability distribution minimizes the output Shannon entropy of the thinning among all the ULC input probability distributions with a given Shannon entropy.
We prove (Theorem \ref{thmthin}) that the geometric distribution minimizes the output entropy of the thinning among all the input probability distributions with a given entropy, without the ad hoc ULC constraint.

Theorem \ref{thmmain} constitutes a strong evidence for the validity of the conjecture in the multimode scenario, whose proof could exploit a multimode generalization of the isoperimetric inequality \eqref{logs}.
The multimode generalization of Theorem \ref{thmmain} is necessary for the proof of the converse theorems for the achievable rates in two communication scenarios.
The first is the triple trade-off coding for simultaneous classical and quantum communication and entanglement sharing, or simultaneous public and private communication and secret key distribution through the Gaussian quantum-limited attenuator \cite{wilde2012public,wilde2012information,wilde2012quantum}.
The second is the transmission of classical information to two receivers through the Gaussian degraded quantum broadcast channel \cite{guha2007classicalproc,guha2007classical}.
The multimode generalization of Theorem \ref{thmmain} would imply the optimality of Gaussian encodings in both scenarios.

The paper is structured as follows.
In Section \ref{setup} we define the Gaussian quantum attenuator and state the main result (Theorem \ref{thmmain}).
Section \ref{seciso} presents the isoperimetric inequality (Theorem \ref{thmiso}).
Theorem \ref{thmmain} is proved in Section \ref{secproof}.
Section \ref{broadcast} discusses the relation with the Gaussian degraded broadcast channel.
Section \ref{secthinning} links the Theorems \ref{thmmain} and \ref{thmiso} to the thinning operation, and the conclusions are in Section \ref{secconcl}.
Finally, Appendix \ref{auxlemmas} contains the proof of some auxiliary lemmas.

\section{Setup and main result}\label{setup}
We consider the Hilbert space of one harmonic oscillator, or one mode of electromagnetic radiation.
Its ladder operator $\hat{a}$ satisfies the canonical commutation relation $\left[\hat{a},\;\hat{a}^\dag\right]=\hat{\mathbb{I}}$, and its Hamiltonian $\hat{N}=\hat{a}^\dag\hat{a}$ counts the number of excitations, or photons.
The state annihilated by $\hat{a}$ is the vacuum $|0\rangle$, from which the Fock states are built:
\begin{equation}\label{fock}
|n\rangle=\frac{\left(\hat{a}^\dag\right)^n}{\sqrt{n!}}|0\rangle\;,\quad\langle m|n\rangle=\delta_{mn}\;,\quad \hat{N}|n\rangle=n|n\rangle\;.
\end{equation}

The quantum-limited attenuator $\Phi_\lambda$ of transmissivity $0\leq\lambda\leq 1$ mixes the input state $\hat{\rho}$ with the vacuum state of an ancillary quantum system $B$ through a beamsplitter of transmissivity $\lambda$.
The beamsplitter is implemented by the unitary operator
\begin{equation}\label{defU}
\hat{U}_\lambda=\exp\left(\left(\hat{a}^\dag\hat{b}-\hat{a}\,\hat{b}^\dag\right)\arccos\sqrt{\lambda}\right)\;,
\end{equation}
that satisfies
\begin{equation}
\hat{U}_\lambda^\dag\;\hat{a}\;\hat{U}_\lambda=\sqrt{\lambda}\;\hat{a}+\sqrt{1-\lambda}\;\hat{b}\;,
\end{equation}
where $\hat{b}$ is the ladder operator of the ancilla system $B$ (see Section 1.4.2 of \cite{ferraro2005gaussian}).
We then have
\begin{equation}
\Phi_\lambda\left(\hat{\rho}\right)=\mathrm{Tr}_B\left[\hat{U}_\lambda\left(\hat{\rho}\otimes |0\rangle_B\langle0|\right)\hat{U}_\lambda^\dag\right]\;.
\end{equation}

For input states diagonal in the Fock basis \eqref{fock}, i.e. with definite photon number, $\Phi_\lambda$ lets each photon be transmitted with probability $\lambda$ and reflected or absorbed with probability $1-\lambda$ (see Section \ref{secthinning}), hence the name ``quantum-limited attenuator''.

The Gaussian thermal state with average energy $E\geq0$ is
\begin{equation}\label{omegaE}
\hat{\omega}_E=\sum_{n=0}^\infty \frac{1}{E+1}\left(\frac{E}{E+1}\right)^n\;|n\rangle\langle n|\;,\quad\mathrm{Tr}\left[\hat{N}\;\hat{\omega}_E\right]=E\;,
\end{equation}
with von Neumann entropy
\begin{equation}\label{defg}
S\left(\hat{\omega}_E\right)=\left(E+1\right)\ln\left(E+1\right)-E\ln E:=g(E)\;,
\end{equation}
and corresponds to a geometric probability distribution of the energy.
The quantum-limited attenuator sends thermal states into themselves, i.e. $\Phi_\lambda\left(\hat{\omega}_E\right)=\hat{\omega}_{\lambda E}$, hence
\begin{equation}
S\left(\Phi_\lambda\left(\hat{\omega}_E\right)\right)=g(\lambda E)=g\left(\lambda\;g^{-1}\left(S\left(\hat{\omega}_E\right)\right)\right)\;.
\end{equation}

Our main result is:
\begin{thm}\label{thmmain}
Gaussian thermal input states \eqref{omegaE} minimize the output entropy of the quantum-limited attenuator among all the input states with a given entropy, i.e. for any input state $\hat{\rho}$ and any $0\leq\lambda\leq1$
\begin{equation}\label{epni}
S\left(\Phi_\lambda\left(\hat{\rho}\right)\right)\geq g\left(\lambda\;g^{-1}\left(S\left(\hat{\rho}\right)\right)\right)\;.
\end{equation}
\end{thm}

\section{Isoperimetric inequality}\label{seciso}
The main step to prove Theorem \ref{thmmain} is the proof of its infinitesimal version.
It states that Gaussian states minimize the derivative of the output entropy of the quantum-limited attenuator with respect to the attenuation parameter for fixed entropy:
\begin{thm}[Isoperimetric inequality]\label{thmiso}
For any quantum state $\hat{\rho}$ with finite support
\begin{equation}\label{logs}
\left.\frac{d}{dt}S\left(\Phi_{e^{-t}}\left(\hat{\rho}\right)\right)\right|_{t=0}:=-F\left(\hat{\rho}\right)\geq f\left(S\left(\hat{\rho}\right)\right)\;,
\end{equation}
where
\begin{equation}\label{deff}
f(S):=-g^{-1}(S)\;g'\left(g^{-1}(S)\right)\;.
\end{equation}
\begin{proof}
The starting point of the proof is the recent result of Ref. \cite{de2015passive}, that links the constrained minimum output entropy conjecture to the notions of passive states.
The passive states of a quantum system \cite{pusz1978passive,lenard1978thermodynamical,gorecki1980passive,vinjanampathy2015quantum,goold2015role,binder2015quantum} minimize the average energy for a given spectrum.
They are diagonal in the energy eigenbasis, and their eigenvalues decrease as the energy increases.
The passive rearrangement $\hat{\rho}^\downarrow$ of a quantum state $\hat{\rho}$ is the only passive state with the same spectrum of $\hat{\rho}$.
The result is the following:
\begin{thm}\label{thmmaj}
The passive rearrangement of the input does not increase the output entropy, i.e. for any quantum state $\hat{\rho}$ and any $0\leq\lambda\leq1$
\begin{equation}\label{passiveS}
S\left(\Phi_\lambda\left(\hat{\rho}\right)\right)\geq S\left(\Phi_\lambda\left(\hat{\rho}^\downarrow\right)\right)\;.
\end{equation}
\begin{proof}
Follows from Theorem 34 and Remark 20 of \cite{de2015passive}.
\end{proof}
\end{thm}
In the following, we will show that this result reduces the proof to the set of passive states.
On this set, we will use the Karush-Kuhn-Tucker (KKT) necessary conditions \cite{kuhn1951} for the maximization of the right-hand side of \eqref{logs} for fixed entropy, and prove that in the limit of infinite support the maximizer tends to a thermal Gaussian state.

Let us fix $S\left(\hat{\rho}\right)=S$.

If $S=0$, from the positivity of the entropy we have for any quantum state $-F\left(\hat{\rho}\right)\geq0=f(0)$,
and the inequality \eqref{logs} is proven.

We can then suppose $S>0$.
Taking the derivative of \eqref{passiveS} with respect to $t$ for $t=0$ we get
\begin{equation}\label{passiveF}
F\left(\hat{\rho}\right)\leq F\left(\hat{\rho}^\downarrow\right)\;,
\end{equation}
hence it is sufficient to prove Theorem \ref{thmiso} for passive states with finite support.

Let us fix $N\in\mathbb{N}$, and consider a quantum state $\hat{\rho}$ with entropy $S$ of the form
\begin{equation}\label{defp}
\hat{\rho}=\sum_{n=0}^N p_n\;|n\rangle\langle n|\;.
\end{equation}
Let $\mathcal{D}_N$ be the set of decreasing probability distributions on $\left\{0,\ldots,\,N\right\}$ with Shannon entropy $S$.
We recall that the Shannon entropy of $p$ coincides with the von Neumann entropy of $\hat{\rho}$.
The state in \eqref{defp} is passive if $p\in\mathcal{D}_N$.

\begin{lem}\label{lemcpt}
$\mathcal{D}_N$ is compact.
\begin{proof}
The set of decreasing probability distributions on $\left\{0,\ldots,\,N\right\}$ is a closed bounded subset of $\mathbb{R}^{N+1}$, hence it is compact.
The Shannon entropy $H$ is continuous on this set.
$\mathcal{D}_N$ is the counterimage of the point $S$, hence it is closed.
Since $\mathcal{D}_N$ is contained in a compact set, it is compact, too.
\end{proof}
\end{lem}

\begin{defn}[Connected support]\label{cons}
A probability distribution $p$ on $\left\{0,\ldots,\,N\right\}$ has connected support iff
$p_n>0$ for $n=0,\ldots,\,N'$, and $p_{N'+1}=\ldots=p_N=0$, where $0\leq N'\leq N$ can depend on $p$
($N'=N$ means $p_n>0$ for any $n$).
We call $\mathcal{P}_N$ the set of probability distributions on $\left\{0,\ldots,\,N\right\}$ with connected support and Shannon entropy $S$.
\end{defn}

We relax the passivity hypothesis, and consider all the states as in \eqref{defp} with $p\in\mathcal{P}_N$.
We notice that any decreasing $p$ has connected support, i.e. $\mathcal{D}_N\subset\mathcal{P}_N$.

From Equations (VIII.5), (VIII.6) and Theorem 56 of Ref. \cite{de2015passive}, we have for any $t\geq0$
\begin{equation}
\Phi_{e^{-t}}\left(\hat{\rho}\right)=\sum_{n=0}^N p_n(t)\;|n\rangle\langle n|\;,
\end{equation}
where
\begin{equation}\label{pnt}
p_n(t)=\sum_{k=n}^N\binom{k}{n}e^{-nt}\left(1-e^{-t}\right)^{k-n}p_k
\end{equation}
satisfies $p_n'(0)=\left(n+1\right)p_{n+1}-n\,p_n$ for $n=0,\ldots,\,N$, and we have set for simplicity $p_{N+1}=0$.

Since $p_{N'+1}=\ldots=p_N=0$, from \eqref{pnt} we get $p_{N'+1}(t)=\ldots=p_N(t)=0$ for any $t\geq0$.
We then have
\begin{equation}
S\left(\Phi_{e^{-t}}\left(\hat{\rho}\right)\right)=-\sum_{n=0}^{N'}p_n(t)\ln p_n(t)\;,
\end{equation}
and
\begin{equation}\label{defF}
F\left(\hat{\rho}\right)=\sum_{n=0}^{N'}p_n'(0)\left(\ln p_n+1\right)=\sum_{n=1}^{N'} n\,p_n\ln\frac{p_{n-1}}{p_n}\;.
\end{equation}

Let $F_N$ be the $\sup$ of $F(p)$ for $p\in\mathcal{P}_N$, where with a bit of abuse of notation we have defined $F(p)=F\left(\hat{\rho}\right)$ for any $\hat{\rho}$ as in \eqref{defp}.
From \eqref{passiveF}, $F_N$ is also the $\sup$ of $F(p)$ for $p\in\mathcal{D}_N$.
From Lemma \ref{lemcpt} $\mathcal{D}_N$ is compact.
Since $F$ is continuous on $\mathcal{D}_N$, the $\sup$ is achieved in a point $p^{(N)}\in\mathcal{D}_N$.
This point satisfies the Karush-Kuhn-Tucker (KKT) necessary conditions \cite{kuhn1951} for the maximization of $F$ with the entropy constraint.
The proof then comes from
\begin{lem}\label{lemmain}
There is a subsequence $\left\{N_k\right\}_{k\in\mathbb{N}}$ such that
\begin{equation}
\lim_{k\to\infty}F_{N_k}=-f(S)\;.
\end{equation}
\begin{proof}
The point $p^{(N)}$ is the maximum of $F$ for $p\in\mathcal{P}_N$.
The constraints read
\begin{equation}
p_0,\ldots,\,p_N\geq0\;,\quad \sum_{n=0}^{N} p_n=1\;,\quad-\sum_{n=0}^{N} p_n\ln p_n=S\;.
\end{equation}
$p^{(N)}$ must then satisfy the associated KKT necessary conditions \cite{kuhn1951}.
We build the functional
\begin{equation}
\tilde{F}(p)=F(p)-\lambda_N\sum_{n=0}^N p_n+\mu_N\sum_{n=0}^N p_n\ln p_n\;.
\end{equation}
Let $N'$ be such that
\begin{equation}\label{pdecr}
p^{(N)}_0\geq\ldots\geq p^{(N)}_{N'}>p^{(N)}_{N'+1}=\ldots=p^{(N)}_N=0\;.
\end{equation}
\begin{rem}\label{remN}
We must have $N'\geq1$.
\end{rem}
Indeed, if $N'=0$ we must have $p_0^{(N)}=1$ and $p^{(N)}_1=\ldots=p^{(N)}_N=0$, hence $S=0$, contradicting the hypothesis $S>0$.

The KKT stationarity condition for $n=0,\ldots,\,N'$ reads
\begin{align}\label{prec}
\left.\frac{\partial}{\partial p_n}\tilde{F}\right|_{p=p^{(N)}} =& \;n\ln\frac{p_{n-1}^{(N)}}{p_n^{(N)}}-n+(n+1)\frac{p_{n+1}^{(N)}}{p_n^{(N)}}\nonumber\\
&-\lambda_N+\mu_N\ln p_n^{(N)}+\mu_N=0\;.
\end{align}
If $N'<N$, $p^{(N)}$ satisfies the KKT dual feasibility condition associated to $p^{(N)}_{N'+1}$.
To avoid the singularity of the logarithm in $0$, we make the variable change
\begin{equation}
y=-p_{N'+1}\ln p_{N'+1}\;,\qquad p_{N'+1}=\psi(y)\;,
\end{equation}
where $\psi$ satisfies
\begin{equation}\label{defpsi}
\psi\left(-x\ln x\right)=x\qquad\forall\;0\leq x\leq\frac{1}{e}\;.
\end{equation}
Since $\psi(0)=0$, the point $p_{N'+1}=0$ corresponds to $y=0$.
Differentiating \eqref{defpsi} with respect to $x$, we get
\begin{equation}
\psi'\left(-x\ln x\right)=-\frac{1}{1+\ln x}\qquad\forall\;0<x<\frac{1}{e}\;,
\end{equation}
and taking the limit for $x\to0$ we get that $\psi'(y)$ is continuous in $y=0$ with $\psi'(0)=0$.

For hypothesis $p^{(N)}\in\mathcal{P}_{N'}\subset\mathcal{P}_{N'+1}\subset\mathcal{P}_N$.
Then, $p^{(N)}$ is a maximum point for $F(p)$ also if we restrict to $p\in\mathcal{P}_{N'+1}$.
We can then consider the restriction of the functional $\tilde{F}$ on $\mathcal{P}_{N'+1}$:
\begin{align}
\tilde{F}(p) =& \sum_{n=1}^{N'}n\,p_n\ln\frac{p_{n-1}}{p_n}+\left(N'+1\right)\psi(y)\ln p_{N'}\nonumber\\
&+\left(N'+1\right)y-{\lambda_N}\sum_{n=0}^{N'}p_n-\lambda_N\;\psi(y)\nonumber\\
&+{\mu_N}\sum_{n=0}^{N'}p_n\ln p_n-\mu_N\,y\;.
\end{align}
The KKT dual feasibility condition is then
\begin{equation}\label{muN}
\left.\frac{\partial}{\partial y}\tilde{F}\right|_{p=p^{(N)}}=N'+1-\mu_N\leq0\;,
\end{equation}
where we have used that $\psi'(0)=0$.

We define for $n=0,\ldots,\,N'$
\begin{equation}
z_n^{(N)}=\frac{p_{n+1}^{(N)}}{p_n^{(N)}}\;.
\end{equation}
Condition \eqref{pdecr} implies
\begin{equation}\label{z01}
0<z_n^{(N)}\leq1\qquad \forall\;n=0,\ldots,\,N'-1\;,\qquad z_{N'}^{(N)}=0\;.
\end{equation}
From Remark \ref{remN} $N'\geq1$, hence $z_0^{(N)}>0$.

Taking the difference of \eqref{prec} for two consecutive values of $n$ we get for $n=0,\ldots,\,N'-1$
\begin{align}\label{znrec}
\left(n+2\right)z_{n+1}^{(N)} =& \left(n+2\right)z_n^{(N)}+1-z_n^{(N)}\nonumber\\
&+\left(1-\mu_N\right)\ln z_n^{(N)}+n\ln\frac{z_n^{(N)}}{z_{n-1}^{(N)}}\;.
\end{align}

\begin{lem}\label{lemincr}
We must have
\begin{equation}\label{mudecr}
1-\mu_N\geq\frac{z_0^{(N)}-1}{\ln z_0^{(N)}}\geq0\;.
\end{equation}
Moreover, $z^{(N)}_n$ is decreasing in $n$ and $N'=N$, i.e.
\begin{equation}
1\geq z^{(N)}_0\geq\ldots\geq z^{(N)}_{N-1}>z^{(N)}_{N}=0\;.
\end{equation}
\begin{proof}
Let us suppose $1-\mu_N<\left.\left(z_0^{(N)}-1\right)\right/\ln z_0^{(N)}$.
We will prove by induction on $n$ that the sequence $z^{(N)}_n$ is increasing in $n$.
The inductive hypothesis is $0<z_0^{(N)}\leq\ldots\leq z_n^{(N)}\leq1$, true for $n=0$.
Since the function $\left.\left(z-1\right)\right/\ln z$ is strictly increasing for $0\leq z\leq1$, we have
\begin{equation}
1-\mu_N<\frac{z_0^{(N)}-1}{\ln z_0^{(N)}}\leq\frac{z_n^{(N)}-1}{\ln z_n^{(N)}}\;,
\end{equation}
and hence $\left(1-\mu_N\right)\ln z_n^{(N)}\geq z_n^{(N)}-1$.
Since $z_{n-1}^{(N)}\leq z_n^{(N)}$, from \eqref{znrec} we have
\begin{eqnarray}
\left(n+2\right)\left(z^{(N)}_{n+1}-z^{(N)}_n\right) &=& 1-z_n^{(N)}+\left(1-\mu_N\right)\ln z_n^{(N)}\nonumber\\
&&+n\ln\frac{z_n^{(N)}}{z_{n-1}^{(N)}}\geq0\;,
\end{eqnarray}
and hence $z_{n+1}^{(N)}\geq z_n^{(N)}$.
However, this is in contradiction with the hypothesis $z^{(N)}_{N'}=0$.

We must then have $1-\mu_N\leq\left.\left(z_0^{(N)}-1\right)\right/\ln z_0^{(N)}$.
We will prove by induction on $n$ that the sequence $z^{(N)}_n$ is decreasing in $n$.
The inductive hypothesis is now $1\geq z_0^{(N)}\geq\ldots\geq z_n^{(N)}>0$, true for $n=0$.
If $n+1=N'$, since $z^{(N)}_{N'}=0$ there is nothing to prove.
We can then suppose $n+1<N'$.
We have
\begin{equation}
1-\mu_N\geq\frac{z_0^{(N)}-1}{\ln z_0^{(N)}}\geq\frac{z_n^{(N)}-1}{\ln z_n^{(N)}}\;,
\end{equation}
and hence $\left(1-\mu_N\right)\ln z_n^{(N)}\leq z_n^{(N)}-1$.
Since $z_{n-1}^{(N)}\geq z_n^{(N)}$, from \eqref{znrec} we have
\begin{eqnarray}
\left(n+2\right)\left(z^{(N)}_{n+1}-z^{(N)}_n\right) &=& 1-z_n^{(N)}+\left(1-\mu_N\right)\ln z_n^{(N)}\nonumber\\
&&+n\ln\frac{z_n^{(N)}}{z_{n-1}^{(N)}}\leq0\;,
\end{eqnarray}
and hence $z_{n+1}^{(N)}\leq z_n^{(N)}$.
Since $n+1< N'$, we also have $z_{n+1}^{(N)}>0$, and the claim is proven.

From Definition \ref{cons} and Remark \ref{remN} we have $1\leq N'\leq N$.
Let us suppose $1\leq N'<N$.
Then, the sequence $p^{(N)}$ satisfies the KKT dual feasibility condition \eqref{muN}, and $N'\leq\mu_N-1$.
From \eqref{mudecr} we get $\mu_N-1\leq0$, hence $N'\leq0$, in contradiction with $N'\geq1$.
We must then have $N'=N$.
\end{proof}
\end{lem}

\begin{lem}\label{lemzbar}
We have
\begin{equation}
\limsup_{N\to\infty}z^{(N)}_{\bar{n}}<1\;,
\end{equation}
where
\begin{equation}
\bar{n}=\min\left\{n\in\mathbb{N}:n+2>e^S\right\}
\end{equation}
does not depend on $N$.
\begin{proof}
We recall that $z_n^{(N)}\leq1$ for any $n$ and $N$, hence
\begin{equation}
\limsup_{N\to\infty}z^{(N)}_{\bar{n}}\leq1\;.
\end{equation}
Let us suppose that $\limsup_{N\to\infty}z^{(N)}_{\bar{n}}=1$.
Then, there is a subsequence $\left\{N_k\right\}_{k\in\mathbb{N}}$ such that $\lim_{k\to\infty}z^{(N_k)}_{\bar{n}}=1$.
Since $z^{(N)}_n$ is decreasing in $n$ for any $N$, we also have
\begin{equation}\label{limbar}
\lim_{k\to\infty}z^{(N_k)}_n=1\qquad \forall\;n=0,\ldots,\,\bar{n}\;.
\end{equation}
Let us define for any $N$ the probability distribution $q^{(N)}\in\mathcal{D}_{\bar{n}+1}$ as
\begin{equation}
q^{(N)}_n=\frac{p^{(N)}_n}{\sum_{k=0}^{\bar{n}+1}p^{(N)}_k}\;,\qquad n=0,\ldots,\,\bar{n}+1\;.
\end{equation}
From \eqref{limbar} we get for $n=0,\ldots,\,\bar{n}+1$
\begin{equation}
\lim_{k\to\infty}\frac{q^{(N_k)}_n}{q^{(N_k)}_0}=\lim_{k\to\infty}z^{(N_k)}_0\ldots z^{(N_k)}_{n-1}=1\;.
\end{equation}
For any $k$
\begin{equation}\label{sumq}
\sum_{n=0}^{\bar{n}+1}q^{(N_k)}_n=1\;.
\end{equation}
Dividing both members of \eqref{sumq} by $q^{(N_k)}_0$ and taking the limit $k\to\infty$ we get $\lim_{k\to\infty}q^{(N_k)}_0=1/\left(\bar{n}+2\right)$, hence $\lim_{k\to\infty}q^{(N_k)}_n=1/\left(\bar{n}+2\right)$ for $n=0,\ldots,\,\bar{n}+1$, and
\begin{equation}
\lim_{k\to\infty}H\left(q^{(N_k)}\right)=\ln\left(\bar{n}+2\right)>S\;.
\end{equation}
However, from Lemma \ref{lemtrunc} we have $H\left(q^{(N)}\right)\leq H\left(p^{(N)}\right)=S$.
\end{proof}
\end{lem}
\begin{cor}\label{corz}
There exists $0\leq\bar{z}<1$ (that does not depend on $N$) such that $z^{(N)}_{\bar{n}}\leq\bar{z}$ for any $N\geq\bar{n}$.
\end{cor}

\begin{lem}
The sequence $\left\{\mu_N\right\}_{N\in\mathbb{N}}$ is bounded.
\begin{proof}
An upper bound for $\mu_N$ is provided by \eqref{mudecr}.
Let us then prove a lower bound.

For any $N\geq\bar{n}+1$ we must have $z_{\bar{n}+1}^{(N)}\geq0$.
The recursive equation \eqref{znrec} for $n=\bar{n}$ gives
\begin{align}\label{znbar}
0 &\leq \left(\bar{n}+2\right)z_{\bar{n}+1}^{(N)}\nonumber\\
&= \left(\bar{n}+1\right)z_{\bar{n}}^{(N)}+1+\left(1-\mu_N\right)\ln z_{\bar{n}}^{(N)}+\bar{n}\ln\frac{z_{\bar{n}}^{(N)}}{z_{\bar{n}-1}^{(N)}}\;.
\end{align}
Since $z^{(N)}_n$ is decreasing in $n$, we have $z_{\bar{n}}^{(N)}\leq z_{\bar{n}-1}^{(N)}$.
Recalling from \eqref{mudecr} that $1-\mu_N\geq0$, and from Corollary \ref{corz} that $z^{(N)}_{\bar{n}}\leq\bar{z}<1$, \eqref{znbar} implies
\begin{eqnarray}
0 &\leq& \left(\bar{n}+1\right)z_{\bar{n}}^{(N)}+1+\left(1-\mu_N\right)\ln z_{\bar{n}}^{(N)}\nonumber\\
&\leq&\left(\bar{n}+1\right)\bar{z}+1+\left(1-\mu_N\right)\ln \bar{z}\;,
\end{eqnarray}
hence $1-\mu_N\leq\left.-\left(\left(\bar{n}+1\right)\bar{z}+1\right)\right/\ln\bar{z}<\infty$.
\end{proof}
\end{lem}
The sequence $\left\{\mu_N\right\}_{N\in\mathbb{N}}$ has then a converging subsequence $\left\{\mu_{N_k}\right\}_{k\in\mathbb{N}}$ with $\lim_{k\to\infty}\mu_{N_k}=\mu$.

Since the sequences $\left\{z^{(N)}_0\right\}_{N\in\mathbb{N}}$ and $\left\{p^{(N)}_0\right\}_{N\in\mathbb{N}}$ are constrained between $0$ and $1$, we can also assume
\begin{equation}
\lim_{k\to\infty}z^{\left(N_k\right)}_0=z_0\;,\qquad\lim_{k\to\infty}p^{\left(N_k\right)}_0=p_0\;.
\end{equation}
Taking the limit of \eqref{mudecr} we get
\begin{equation}\label{mulim0}
1-\mu\geq\frac{z_0-1}{\ln z_0}\geq0\;.
\end{equation}

\begin{lem}\label{lemzn}
$\lim_{k\to\infty}z^{\left(N_k\right)}_n=z_n$ for any $n\in\mathbb{N}$,
where the $z_n$ are either all $0$ or all strictly positive, and in the latter case they satisfy for any $n$ in $\mathbb{N}$ the recursive relation \eqref{znrec} with $\mu_N$ replaced by $\mu$:
\begin{eqnarray}\label{zreclim}
\left(n+2\right)z_{n+1} &=& \left(n+2\right)z_n+1-z_n+\left(1-\mu\right)\ln z_n\nonumber\\
&&+n\ln\frac{z_n}{z_{n-1}}\;.
\end{eqnarray}
\begin{proof}
If $z_0=0$, since $z^{(N)}_n$ is decreasing in $n$ we have for any $n$ in $\mathbb{N}$
\begin{equation}
\limsup_{k\to\infty}z^{(N_k)}_n\leq\limsup_{k\to\infty}z^{(N_k)}_0=z_0=0\;,
\end{equation}
hence $\lim_{k\to\infty}z^{(N_k)}_n=0$.

Let us now suppose $z_0>0$, and proceed by induction on $n$.
From the inductive hypothesis, we can suppose
\begin{equation}
z_0=\lim_{k\to\infty}z^{\left(N_k\right)}_0\geq\ldots\geq\lim_{k\to\infty}z^{\left(N_k\right)}_n=z_n>0\;.
\end{equation}
Then, taking the limit in \eqref{znrec} we get
\begin{align}
z_{n+1} &= \lim_{k\to\infty}z^{\left(N_k\right)}_{n+1}\nonumber\\
&=z_n+\frac{1-z_n+\left(1-\mu\right)\ln z_n+n\ln\frac{z_n}{z_{n-1}}}{n+2}\;.
\end{align}
If $z_{n+1}>0$, the claim is proven.
Let us then suppose $z_{n+1}=0$.
From \eqref{znrec} we get then
\begin{equation}
0\leq\lim_{k\to\infty}z^{\left(N_k\right)}_{n+2}=\frac{1+\left(n+2-\mu\right)\ln0-\left(n+1\right)\ln z_n}{n+3}\;,
\end{equation}
that implies $\mu\geq n+2\geq2$.
However, \eqref{mulim0} implies $\mu\leq1$.
\end{proof}
\end{lem}

\begin{lem}\label{zlim}
There exists $0\leq z<1$ such that $z_n=\lim_{k\to\infty}z^{\left(N_k\right)}_n=z$ for any $n\in\mathbb{N}$.
\begin{proof}
If $z_0=0$, Lemma \ref{lemzn} implies the claim with $z=0$.
Let us then suppose $z_0>0$.

If $z_0=1$, we can prove by induction on $n$ that $z_n=1$ for any $n\in\mathbb{N}$.
The claim is true for $n=0$.
From the inductive hypothesis we can suppose $z_0=\ldots=z_n=1$.
The relation \eqref{zreclim} implies then $z_{n+1}=1$.

However, from Lemma \ref{lemzn} and Corollary \ref{corz} we must have $z_{\bar{n}}=\lim_{k\to\infty}z^{\left(N_k\right)}_{\bar{n}}\leq\bar{z}<1$.
Then, it must be $0<z_0<1$.

Since the sequence $\left\{z^{(N)}_n\right\}_{n\in\mathbb{N}}$ is decreasing for any $N$, also the sequence $\left\{z_n\right\}_{n\in\mathbb{N}}$ is decreasing.
Since it is also positive, it has a limit $\lim_{n\to\infty}z_n=\inf_{n\in\mathbb{N}}z_n=z$, that satisfies $0\leq z\leq z_0<1$.
Since $z_n\leq z_{n-1}\leq z_0<1$, \eqref{zreclim} implies
\begin{equation}
\left(n+2\right)\left(z_n-z_{n+1}\right)+1-z_n+\left(1-\mu\right)\ln z_n\geq0\;,
\end{equation}
hence
\begin{equation}\label{zdecrineq}
1-\mu\leq\frac{\left(n+2\right)\left(z_n-z_{n+1}\right)}{-\ln z_n}+\frac{z_n-1}{\ln z_n}\;.
\end{equation}
The sequence $\left\{z_n-z_{n+1}\right\}_{n\in\mathbb{N}}$ is positive and satisfies
\begin{equation}
\sum_{n=0}^\infty\left(z_n-z_{n+1}\right)=z_0-z<\infty\;.
\end{equation}
Then, from Lemma \ref{lemnx} $\liminf_{n\to\infty}\left(n+2\right)\left(z_n-z_{n+1}\right)=0$, and since $-\ln z_n\geq-\ln z_0>0$, also
\begin{equation}
\liminf_{n\to\infty}\frac{\left(n+2\right)\left(z_n-z_{n+1}\right)}{-\ln z_n}=0\;.
\end{equation}
Then, taking the $\liminf$ of \eqref{zdecrineq} for $n\to\infty$ we get $1-\mu\leq\left.\left(z-1\right)\right/\ln z$.
Combining with \eqref{mulim0} and recalling that $z\leq z_0$ we get
\begin{equation}\label{mulim}
\frac{z-1}{\ln z}\leq\frac{z_0-1}{\ln z_0}\leq1-\mu\leq\frac{z-1}{\ln z}\;,
\end{equation}
that implies $z=z_0$.
Since $z_n$ is decreasing and $z=\inf_{n\in\mathbb{N}}z_n$, we have $z_0=z\leq z_n\leq z_0$ for any $n$, hence $z_n=z$.
\end{proof}
\end{lem}

\begin{lem}\label{lemp0lim}
$\lim_{k\to\infty}p^{\left(N_k\right)}_n=p_0\,z^n$ for any $n\in\mathbb{N}$.
\begin{proof}
The claim is true for $n=0$.
The inductive hypothesis is $\lim_{k\to\infty}p^{\left(N_k\right)}_{n'}=p_0\,z^{n'}$ for $n'=0,\ldots,\,n$.
We then have $\lim_{k\to\infty}p^{\left(N_k\right)}_{n+1}=\lim_{k\to\infty}p^{\left(N_k\right)}_n\,z^{\left(N_k\right)}_n=p_0\,z^{n+1}$,
where we have used the inductive hypothesis and Lemma \ref{zlim}.
\end{proof}
\end{lem}
\begin{lem}\label{limp}
$p_0=1-z$, hence $\lim_{k\to\infty}p^{\left(N_k\right)}_n=\left(1-z\right)z^n$ for any $n\in\mathbb{N}$.
\begin{proof}
We have $\sum_{n=0}^N p^{(N)}_n=1$ for any $N\in\mathbb{N}$.
Moreover, since $z^{(N)}_n$ is decreasing in $n$, we also have
\begin{equation}
p^{(N)}_n=p^{(N)}_0\,z^{(N)}_0\ldots\,z^{(N)}_{n-1}\leq p^{(N)}_0\left(z^{(N)}_0\right)^n\;.
\end{equation}
Since $\lim_{k\to\infty}z_0^{\left(N_k\right)}=z<1$, for sufficiently large $k$ we have $z_0^{\left(N_k\right)}\leq\left(1+z\right)/2$, and since $p^{(N)}_0\leq1$,
\begin{equation}\label{boundp}
p^{\left(N_k\right)}_n\leq\left(\frac{1+z}{2}\right)^n\;.
\end{equation}
The sums $\sum_{n=0}^{N_k} p^{\left(N_k\right)}_n$ are then dominated for any $k\in\mathbb{N}$ by $\sum_{n=0}^\infty \left(\frac{1+z}{2}\right)^n<\infty$, and from the dominated convergence theorem we have
\begin{eqnarray}
1 &=& \lim_{k\to\infty}\sum_{n=0}^{N_k}p^{\left(N_k\right)}_n=\sum_{n=0}^\infty\lim_{k\to\infty}p^{\left(N_k\right)}_n=p_0\sum_{n=0}^\infty z^n\nonumber\\
&=&\frac{p_0}{1-z}\;,
\end{eqnarray}
where we have used Lemma \ref{lemp0lim}.
\end{proof}
\end{lem}

\begin{lem}\label{z(S)}
$z=g^{-1}(S)\left/\left(g^{-1}(S)+1\right)\right.$.
\begin{proof}
The function $-x\ln x$ is increasing for $0\leq x\leq1/e$.
Let us choose $n_0$ such that $\left(\left.\left(1+z\right)\right/2\right)^{n_0}\leq1/e$.
Recalling \eqref{boundp}, the sums $-\sum_{n=n_0}^{N_k} p^{(N_k)}_n\ln p^{(N_k)}_n$ are dominated for any $k\in\mathbb{N}$ by $-\sum_{n=n_0}^\infty n\left(\frac{1+z}{2}\right)^n\ln\frac{1+z}{2}<\infty$.
For any $N$ we have $S=-\sum_{n=0}^N p^{(N)}_n\ln p^{(N)}_n$.
Then, from the dominated convergence theorem and Lemma \ref{limp} we have
\begin{align}\label{Sz}
S &= -\sum_{n=0}^\infty\lim_{k\to\infty}p^{(N_k)}_n\ln p^{(N_k)}_n\nonumber\\
&= -\sum_{n=0}^\infty\left(1-z\right)z^n\left(\ln\left(1-z\right)+n\ln z\right)=g\left(\frac{z}{1-z}\right)\;,
\end{align}
where we have used the definition of $g$ \eqref{defg}.
Finally, the claim follows solving \eqref{Sz} with respect to $z$.
\end{proof}
\end{lem}

It is convenient to rewrite $F_{N_k}=F\left(p^{(N_k)}\right)$ as
\begin{equation}\label{FN}
F_{N_k}=-\sum_{n=0}^{N_k-1}\left(n+1\right)p^{(N_k)}_n\,z^{(N_k)}_n\ln z^{(N_k)}_n\;.
\end{equation}
Since $z^{(N_k)}_n\leq1$, each term of the sum is positive.
Since $-x\ln x\leq1/e$ for $0\leq x\leq1$, and recalling \eqref{boundp}, the sum is dominated by $\sum_{n=0}^\infty \frac{n+1}{e}\left(\frac{1+z}{2}\right)^n<\infty$.
We then have from the dominated convergence theorem, recalling Lemmas \ref{limp} and \ref{zlim},
\begin{align}
\lim_{k\to\infty}F_{N_k} &= -\sum_{n=0}^\infty\left(n+1\right)\lim_{k\to\infty}p^{(N_k)}_n\,z^{(N_k)}_n\ln z^{(N_k)}_n=\nonumber\\
&= -\sum_{n=0}^\infty\left(n+1\right)\left(1-z\right)z^{n+1}\ln z=\frac{z\ln z}{z-1}=\nonumber\\
&= g^{-1}(S)\ln\left(1+\frac{1}{g^{-1}(S)}\right)=-f(S)\;,
\end{align}
where we have used Lemma \ref{z(S)} and the definitions of $f$ \eqref{deff} and $g$ \eqref{defg}.
\end{proof}
\end{lem}

Then, since $\mathcal{D}_N\subset\mathcal{D}_{N+1}$ for any $N$, $F_N$ is increasing in $N$, and for any $p\in\mathcal{P}_N$
\begin{equation}
F(p)\leq F_N\leq \sup_{N\in\mathbb{N}}F_N=\lim_{N\to\infty}F_N=\lim_{k\to\infty}F_{N_k}=-f(S).
\end{equation}
\end{proof}
\end{thm}

\section{Proof of Theorem \ref{thmmain}}\label{secproof}
The idea for the proof is integrating the infinitesimal version \eqref{logs}.

From Theorem \ref{thmmaj}, it is sufficient to prove Theorem \ref{thmmain} for passive states, i.e. states of the form
\begin{equation}
\hat{\rho}=\sum_{n=0}^\infty p_n\;|n\rangle\langle n|\;,\qquad p_0\geq p_1\geq\ldots\geq0\;.
\end{equation}

\begin{lem}\label{finites}
If Theorem \ref{thmmain} holds for any passive state with finite support, then it holds for any passive state.
\begin{proof}
Let $\hat{\rho}$ be a passive state.
If $S\left(\Phi_\lambda\left(\hat{\rho}\right)\right)=\infty$, there is nothing to prove.
We can then suppose $S\left(\Phi_\lambda\left(\hat{\rho}\right)\right)<\infty$.

We can associate to $\hat{\rho}$ the probability distribution $p$ on $\mathbb{N}$ such that
\begin{equation}
\hat{\rho}=\sum_{n=0}^\infty p_n\;|n\rangle\langle n|\;,
\end{equation}
satisfying $-\sum_{n=0}^\infty p_n\ln p_n=S\left(\hat{\rho}\right)$.
Let us define for any $N\in\mathbb{N}$ the quantum state
\begin{equation}
\hat{\rho}_N=\sum_{n=0}^N \frac{p_n}{s_N}\;|n\rangle\langle n|\;,
\end{equation}
where $s_N=\sum_{n=0}^N p_n$.
We have
\begin{equation}
\left\|\hat{\rho}_N-\hat{\rho}\right\|_1 = \frac{1-s_N}{s_N}\sum_{n=0}^Np_n+\sum_{n=N+1}^\infty p_n\;,
\end{equation}
where $\left\|\cdot\right\|_1$ denotes the trace norm \cite{wilde2013quantum,holevo2013quantum}.
Since
\begin{equation}
\lim_{N\to\infty}s_N=1\;,\qquad \sum_{n=0}^\infty p_n=1\;,
\end{equation}
we have $\lim_{N\to\infty}\left\|\hat{\rho}_N-\hat{\rho}\right\|_1 =0$.
Since $\Phi_\lambda$ is continuous in the trace norm, we also have
\begin{equation}\label{limPhiN}
\lim_{N\to\infty}\left\|\Phi_\lambda\left(\hat{\rho}_N\right)-\Phi_\lambda\left(\hat{\rho}\right)\right\|_1=0\;.
\end{equation}
Moreover,
\begin{equation}\label{limSN}
\lim_{N\to\infty}S\left(\hat{\rho}_N\right)=\lim_{N\to\infty}\left(\ln s_N-\sum_{n=0}^N \frac{p_n}{s_N}\ln p_n\right)=S\left(\hat{\rho}\right)\;.
\end{equation}
Notice that \eqref{limSN} holds also if $S\left(\hat{\rho}\right)=\infty$.

Let us now define the probability distribution $q$ on $\mathbb{N}$ as
\begin{equation}
\Phi_\lambda\left(\hat{\rho}\right)=\sum_{n=0}^\infty q_n\;|n\rangle\langle n|\;,
\end{equation}
satisfying
\begin{equation}\label{Sq}
S\left(\Phi_\lambda\left(\hat{\rho}\right)\right)=-\sum_{n=0}^\infty q_n\ln q_n\;.
\end{equation}
From \cite{ivan2011operator}, Section IV.B, or \cite{de2015passive}, Equation (II.12), the channel $\Phi_\lambda$ sends the set of states supported on the span of the first $N+1$ Fock states into itself.
Then, for any $N\in\mathbb{N}$ there is a probability distribution $q^{(N)}$ on $\left\{0,\ldots,\,N\right\}$ such that
\begin{equation}
\Phi_\lambda\left(\hat{\rho}_N\right)=\sum_{n=0}^N q_n^{(N)}\;|n\rangle\langle n|\;.
\end{equation}
From \eqref{limPhiN} we get for any $n\in\mathbb{N}$
\begin{equation}\label{limqN}
\lim_{N\to\infty}q^{(N)}_n=q_n\;.
\end{equation}
Since $\Phi_\lambda$ is trace preserving, we have $\sum_{n=0}^\infty q_n=1$, hence $\lim_{n\to\infty} q_n=0$.
Then, there is $n_0\in\mathbb{N}$ (that does not depend on $N$) such that for any $n\geq n_0$ we have $q_n\leq p_0/e$.
Since $s_N\;\hat{\rho}_N\leq\hat{\rho}$ and the channel $\Phi_\lambda$ is positive, we have $s_N\;\Phi_\lambda\left(\hat{\rho}_N\right)\leq\Phi_\lambda\left(\hat{\rho}\right)$.
Then, for any $n\geq n_0$
\begin{equation}
q_n^{(N)}\leq\frac{q_n}{s_N}\leq\frac{q_n}{p_0}\leq\frac{1}{e}\;,
\end{equation}
where we have used that $s_N\geq p_0>0$.
Since the function $-x\ln x$ is increasing for $0\leq x\leq1/e$, the sums
\begin{equation}
-\sum_{n=n_0}^N q^{(N)}_n\ln q^{(N)}_n
\end{equation}
are dominated by
\begin{equation}
\sum_{n=n_0}^\infty \frac{q_n\ln p_0-q_n\ln q_n}{p_0}\leq\frac{\ln p_0+S\left(\Phi_\lambda\left(\hat{\rho}\right)\right)}{p_0}<\infty\;,
\end{equation}
where we have used \eqref{Sq}.
Then, from the dominated convergence theorem we have
\begin{eqnarray}
\lim_{N\to\infty}S\left(\Phi_\lambda\left(\hat{\rho}_N\right)\right) &=& -\lim_{N\to\infty}\sum_{n=0}^N q^{(N)}_n\ln q^{(N)}_n\nonumber\\
&=& -\sum_{n=0}^\infty\lim_{N\to\infty}q^{(N)}_n\ln q^{(N)}_n\nonumber\\
&=& S\left(\Phi_\lambda\left(\hat{\rho}\right)\right)\;,
\end{eqnarray}
where we have also used \eqref{limqN}.

If Theorem \ref{thmmain} holds for passive states with finite support, for any $N$ in $\mathbb{N}$ we have
\begin{equation}
S\left(\Phi_\lambda\left(\hat{\rho}_N\right)\right)\geq g\left(\lambda\;g^{-1}\left(S\left(\hat{\rho}_N\right)\right)\right)\;.
\end{equation}
Then, the claim follows taking the limit $N\to\infty$.
\end{proof}
\end{lem}
From Lemma \ref{finites}, we can suppose $\hat{\rho}$ to be a passive state with finite support.
\begin{lem}\label{lemcomp}
The quantum-limited attenuator $\Phi_\lambda$ satisfies the composition rule $\Phi_\lambda\circ\Phi_{\lambda'}=\Phi_{\lambda\,\lambda'}$.
\begin{proof}
Follows from Lemma 13 of \cite{de2015passive}.
\end{proof}
\end{lem}
The function $g(x)$ defined in \eqref{defg} is differentiable for $x>0$, and continuous and strictly increasing for $x\geq0$, and its image is the whole interval $[0,\,\infty)$.
Then, its inverse $g^{-1}(S)$ is defined for any $S\geq0$, it is continuous and strictly increasing for $S\geq0$, and differentiable for $S>0$.
We define for any $t\geq0$ the functions
\begin{equation}
\phi(t)=S\left(\Phi_{e^{-t}}\left(\hat{\rho}\right)\right)\;,\qquad\phi_0(t)=g\left(e^{-t}\;g^{-1}\left(S\left(\hat{\rho}\right)\right)\right)\;.
\end{equation}
It is easy to show that
\begin{equation}\label{start}
\phi(0)=\phi_0(0)\;,
\end{equation}
and
\begin{equation}\label{phi0}
\frac{d}{dt}\phi_0(t)=f\left(\phi_0(t)\right)\;,
\end{equation}
where $f$ is defined by \eqref{deff}.
\begin{lem}
$f$ is differentiable for any $S\geq0$.
\begin{proof}
We have
\begin{equation}
f'(S)=\frac{1}{\left(1+g^{-1}(S)\right)\ln\left(1+\frac{1}{g^{-1}(S)}\right)}-1\;,
\end{equation}
hence $\lim_{S\to0}f'(S)=-1$.
\end{proof}
\end{lem}

Since the quantum-limited attenuator sends the set of passive states with finite support into itself (see Equation (II.12) of \cite{de2015passive}), we can replace $\hat{\rho}\to \Phi_{e^{-t}}\left(\hat{\rho}\right)$ in equation \eqref{logs}, and from Theorem \ref{thmiso} and Lemma \ref{lemcomp} we get
\begin{equation}\label{phi}
\frac{d}{dt}\phi(t)\geq f\left(\phi(t)\right)\;.
\end{equation}
The claim then follows from
\begin{thm}[Comparison theorem for first-order ordinary differential equations]
Let $\phi,\,\phi_0:[0,\infty)\to[0,\infty)$ be differentiable functions satisfying \eqref{start}, \eqref{phi0} and \eqref{phi} with $f:[0,\infty)\to\mathbb{R}$ differentiable.
Then, $\phi(t)\geq\phi_0(t)$ for any $t\geq0$.
\begin{proof}
See e.g. Theorem 2.2.2 of \cite{ames1997inequalities}.
\end{proof}
\end{thm}

\section{Relation with the degraded Gaussian broadcast channel}\label{broadcast}
The quantum degraded Gaussian broadcast channel \cite{guha2007classicalproc,guha2007classical} maps a state $\hat{\rho}_A$ of the quantum system $A$ to a state $\hat{\rho}_{A'B'}$ of the joint quantum system $A'B'$ with
\begin{equation}\label{defbr}
\hat{\rho}_{A'B'}=\hat{U}_\lambda\left(\hat{\rho}_A\otimes|0\rangle_B\langle0|\right)\hat{U}_\lambda^\dag\;,
\end{equation}
where $\hat{U}_\lambda$ is the unitary operator defined in \eqref{defU}, and $1/2\leq\lambda\leq1$.
It can be understood as follows.
$A$ encodes the information into the state of the electromagnetic radiation $\hat{\rho}_A$, and sends it through a beamsplitter of transmissivity $\lambda$.
$A'$ and $B'$ receive the transmitted and the reflected part of the signal, respectively, whose joint state is $\hat{\rho}_{A'B'}$.

The rate pair $\left(R_{A'},\,R_{B'}\right)$ is achievable if for any $n\in\mathbb{N}$ $A$ can send to $A'$ and $B'$ with $n$ uses of the channel any couple of messages chosen from sets $I_{A'}^{(n)}$ and $I_{B'}^{(n)}$ with
\begin{equation}
\left|I_{A'}^{(n)}\right|\geq e^{nR_{A'}}\;,\qquad\left|I_{B'}^{(n)}\right|\geq e^{nR_{B'}}
\end{equation}
with vanishing maximum error probability in the limit $n\to\infty$ (see \cite{yard2011quantum}, Section II, and \cite{guha2007classical}, Sections II and III; see also \cite{savov2015classical}, Definition 1).
The closure of the set of the achievable rate pairs constitutes the capacity region of the channel.

Let $E>0$ be the maximum average energy per copy of the input states.
Ref. \cite{guha2007classical} first proves in Section IV that superposition coding allows to achieve with the quantum degraded Gaussian broadcast channel \eqref{defbr} any rate pair $(R_{A'},\,R_{B'})$ satisfying
\begin{align}\label{RA2}
R_{A'} &\leq g\left(\lambda\,\beta\,E\right)\;,\\
R_{B'} &\leq g\left(\left(1-\lambda\right)E\right)-g\left(\left(1-\lambda\right)\beta\,E\right)\label{RB2}
\end{align}
for some $0\leq\beta\leq1$.
For the converse, Ref. \cite{guha2007classical} proves an outer bound for the achievable rate pairs in Eqs. (22) and (23).
Any achievable rate pair $\left(R_{A'},\,R_{B'}\right)$ must satisfy the following property.
For any $n\in\mathbb{N}$ there must exist an ensemble of encoding \emph{pure} states $\left\{p^{(n)}_iq^{(n)}_j,\;\hat{\rho}^{A(n)}_{ij}\right\}$ on $n$ copies of the quantum system $A$ such that 
\begin{align}\label{RA}
n\,R_{A'} \leq \sum_j q^{(n)}_j & \left(S\left(\hat{\rho}^{A'(n)}_j\right)\phantom{\sum_i}\right.\nonumber\\
&\left.\;\;-\sum_ip^{(n)}_i\;S\left(\hat{\rho}^{A'(n)}_{ij}\right)\right) + n\,\epsilon_n''\;,
\end{align}
\begin{equation}\label{RB}
n\,R_{B'} \leq S\left(\hat{\rho}^{(n)}_{B'}\right)-\sum_jq^{(n)}_j\;S\left(\hat{\rho}^{B'(n)}_j\right) + n\,\epsilon_n'\;,
\end{equation}
where $\epsilon_n',\,\epsilon_n''\to0$ for $n\to\infty$, and
\begin{align}
\hat{\rho}^{A'B'(n)}_{ij} &= \hat{U}_\lambda^{\otimes n}\left(\hat{\rho}^{A(n)}_{ij}\otimes\left(|0\rangle_B\langle0|\right)^{\otimes n}\right)\hat{U}_\lambda^{\dag\otimes n}\;,\\
\hat{\rho}^{B'(n)}_j &= \sum_i p^{(n)}_i\;\hat{\rho}^{B'(n)}_{ij}\;,\label{rhoBj}\\
\hat{\rho}^{(n)}_{B'} &= \sum_j q^{(n)}_j\;\hat{\rho}^{B'(n)}_j\;.
\end{align}
Using the outer bounds \eqref{RA}, \eqref{RB} and assuming the multimode version of the inequality \eqref{epni}, Ref. \cite{guha2007classical} proves that any achievable rate pair $(R_{A'},R_{B'})$ must satisfy \eqref{RA2} and \eqref{RB2} for some $0\leq\beta\leq1$.
Eqs. \eqref{RA2} and \eqref{RB2} then describe the capacity region of the quantum degraded Gaussian broadcast channel.

One may ask whether the one-mode inequality \eqref{epni} implies the outer bounds \eqref{RA2}, \eqref{RB2} in the setting where the sender $A$ cannot entangle the input state among successive uses of the channel, i.e. when the pure states $\hat{\rho}^{A(n)}_{ij}$ are product states.
This would be the case if the bounds \eqref{RA}, \eqref{RB} were additive, i.e. if they did not require the regularization over $n$.
In this case determining them for $n=1$ would be sufficient.
The answer is negative.
Indeed, the subadditivity of the entropy for the terms $S\left(\hat{\rho}^{B'(n)}_j\right)$ in \eqref{RB} goes in the wrong direction.
Additivity would hold if $\hat{\rho}^{B'(n)}_j$ were product states, but from \eqref{rhoBj} in general this is not the case.

\section{The thinning}\label{secthinning}
The thinning \cite{renyi1956characterization} is the map acting on classical probability distributions on the set of natural numbers that is the discrete analogue of the continuous rescaling operation on positive real numbers.
\begin{defn}[Thinning]
Let $N$ be a random variable with values in $\mathbb{N}$.
The thinning with parameter $0\leq\lambda\leq1$ is defined as
\begin{equation}
T_\lambda(N)=\sum_{i=1}^N B_i\;,
\end{equation}
where the $\{B_n\}_{n\in\mathbb{N}^+}$ are independent Bernoulli variables with parameter $\lambda$, i.e. each $B_i$ is one with probability $\lambda$, and zero with probability $1-\lambda$.
\end{defn}
From a physical point of view, the thinning can be understood as follows: each incoming photon has probability $\lambda$ of being transmitted, and $1-\lambda$ of being reflected or absorbed.
Let $N$ be the random variable associated to the number of incoming photons, and $\{p_n\}_{n\in\mathbb{N}}$ its probability distribution, i.e. $p_n$ is the probability that $N=n$ (i.e. that $n$ photons are sent).
Then, $T_\lambda(p)$ is the probability distribution of the number of transmitted photons.
It is easy to show that
\begin{equation}\label{Tn}
\left[T_\lambda(p)\right]_n=\sum_{k=0}^\infty r_{n|k}\;p_k\;,
\end{equation}
where the transition probabilities $r_{n|k}$ are given by
\begin{equation}\label{rnk}
r_{n|k}=\binom{k}{n}\lambda^n(1-\lambda)^{k-n}\;,
\end{equation}
and vanish for $k<n$.

The thinning coincides with the restriction of the attenuator to input states diagonal in the Fock basis:
\begin{thm}\label{thinatt}
Let $\Phi_\lambda$ and $T_\lambda$ be the quantum-limited attenuator and the thinning of parameter $0\leq\lambda\leq1$, respectively.
Then for any probability distribution $p$ on $\mathbb{N}$
\begin{equation}
\Phi_\lambda\left(\sum_{n=0}^\infty p_n\;|n\rangle\langle n|\right)=\sum_{n=0}^\infty \left[T_\lambda(p)\right]_n\;|n\rangle\langle n|\;.
\end{equation}
\begin{proof}
See Theorem 56 of \cite{de2015passive}.
\end{proof}
\end{thm}
Thanks to Theorem \ref{thinatt}, our main results Theorems \ref{thmmain} and \ref{thmiso} apply also to the thinning:
\begin{thm}\label{thmthin}
Geometric input probability distributions minimize the output Shannon entropy of the thinning for fixed input entropy, i.e. for any probability distribution $p$ on $\mathbb{N}$ and any $0\leq\lambda\leq 1$ we have
\begin{equation}
H\left(T_\lambda(p)\right)\geq g\left(\lambda\;g^{-1}\left(H(p)\right)\right)\;,
\end{equation}
where $g$ has been defined in \eqref{defg}.
\end{thm}
\begin{thm}
For any probability distribution $p$ on $\mathbb{N}$
\begin{equation}
\left.\frac{d}{dt}H\left(T_{e^{-t}}(p)\right)\right|_{t=0}\geq f\left(H(p)\right)\;,
\end{equation}
where $f$ has been defined in \eqref{deff}.
\end{thm}

\section{Conclusion}\label{secconcl}
We have proved that Gaussian thermal input states minimize the output von Neumann entropy of the Gaussian quantum-limited attenuator for fixed input entropy (Theorem \ref{thmmain}).
The proof is based on a new isoperimetric inequality (Theorem \ref{thmiso}).
Theorem \ref{thmmain} implies that geometric input probability distributions minimize the output Shannon entropy of the thinning for fixed input entropy (Theorem \ref{thmthin}).
Its multimode extension would permit to determine both the triple trade-off region of the Gaussian quantum-limited attenuator \cite{wilde2012public,wilde2012information,wilde2012quantum} and the classical capacity region of the Gaussian quantum degraded broadcast channel \cite{guha2007classicalproc,guha2007classical}.
The proof of Theorem \ref{thmmain} for the multimode attenuator would follow from the multimode generalization of the isoperimetric inequality \eqref{logs}.
However, our proof of \eqref{logs} relies on the majorization result of Ref. \cite{de2015passive}, that does not hold for the multimode attenuator (see \cite{de2016passive}, Section IV.A).

\appendices
\section{Auxiliary Lemmas}\label{auxlemmas}
\begin{lem}\label{lemtrunc}
Let us choose a probability distribution $p\in\mathcal{D}_N$, fix $0\leq N'\leq N$, and define the probability distribution $q\in\mathcal{D}_{N'}$ as
\begin{equation}
q_n=\frac{p_n}{\sum_{k=0}^{N'}p_k}\;,\qquad n=0,\ldots,\,N'\;.
\end{equation}
Then, $H(q)\leq H(p)$.
\begin{proof}
We have for $n=0,\ldots,\,N'$
\begin{equation}
\sum_{k=0}^n q_k=\frac{\sum_{k=0}^n p_k}{\sum_{l=0}^{N'}p_l}\geq\sum_{k=0}^n p_k\;,
\end{equation}
Then, $q$ majorizes $p$ and the claim follows from Remark 20 of \cite{de2015passive}.
\end{proof}
\end{lem}

\begin{lem}\label{lemnx}
Let $\left\{x_n\right\}_{n\in\mathbb{N}}$ be a positive sequence with finite sum.
Then $\liminf_{n\to\infty}n\,x_n=0$.
\begin{proof}
Let us suppose $\liminf_{n\to\infty}n\,x_n=c>0$.
Then, there exists $n_0\in\mathbb{N}$ such that $n\,x_n\geq c/2$ for any $n\geq n_0$.
Then,
\begin{equation}
\sum_{n=0}^\infty x_n\geq\sum_{n=n_0}^\infty\frac{c}{2n}=\infty\;,
\end{equation}
contradicting the hypothesis.
\end{proof}
\end{lem}

\section*{Acknowledgment}
The Authors  thank Luigi Ambrosio for comments and fruitful discussions.

\bibliography{biblio}
\bibliographystyle{IEEEtran}

\begin{IEEEbiographynophoto}{Giacomo De Palma}
was born in Lanciano (CH), Italy, on March 15, 1990.
He received the B.S. degree in Physics and the M.S. degree in Physics from the University of Pisa, Pisa (PI), Italy, in 2011 and 2013, respectively. He also received the ``Diploma di Licenza'' in Physics and the Ph.D. degree in Physics from Scuola Normale Superiore, Pisa (PI), Italy, in 2014 and 2016, respectively.

He is currently a postdoc at the University of Copenhagen, Copenhagen, Denmark.

His research interests include quantum information, quantum statistical mechanics and quantum thermodynamics.
He is author of eleven scientific papers published in peer-reviewed journals.
\end{IEEEbiographynophoto}
\begin{IEEEbiographynophoto}{Dario Trevisan}
was born in the Province of Venice, Italy. He received the B.S. degree in mathematics from the University of Pisa, Pisa, Italy, in 2009, the M.S. degree in mathematics from the University of Pisa, in 2011, and the Ph.D. degree in mathematics from the Scuola Normale Superiore, Pisa, Italy, in 2014.

He is currently Assistant Professor at the University of Pisa.

Dr. Trevisan is a member of the GNAMPA group of the Italian National Institute for Higher Mathematics (INdAM).
\end{IEEEbiographynophoto}
\begin{IEEEbiographynophoto}{Vittorio Giovannetti}
was born in Castelnuovo di Garfagnana (LU) Italy, on April 1, 1970. He received the M.S. degree in Physics from the University of Pisa and PhD degree in theoretical Physics from the University of Perugia.

He is currently Associate Professor at the Scuola Normale Superiore of Pisa.
\end{IEEEbiographynophoto}
\end{document}